\documentclass[preprint,fleqn,pre,showpacs,onecolumn,12pt]{revtex4}%
\usepackage{amssymb}
\usepackage{amsmath}
\usepackage{graphicx}
\usepackage{amsfonts}%
\begin{document}
\title{Pattern formation in parametric sound generation}
\author{Isabel P\'{e}rez-Arjona and V\'{\i}ctor J. S\'{a}nchez-Morcillo }
\affiliation{Departament de F\'{\i}sica Aplicada, Universitat Polit\'{e}cnica de
Val\`{e}ncia, Crta. Natzaret-Oliva s/n, 46730 Grau de Gandia, Spain}
\author{Short title: Pattern formation in acoustics}
\begin{abstract}
Pattern formation of sound is predicted in a driven resonator where
subharmonic generation takes place. A model allowing for diffraction of the
fields (large-aspect ratio limit) is derived by means of the multiple scale
expansions technique. An analysis of the solutions and its stability against
space-dependent perturbations is performed in detail considering the
distinctive peculiarities of the acoustical system. Numerical integration
confirm the analytical predictions, and shows the possibility of patterns in
the form of stripes and squares.

\end{abstract}

\maketitle

\section{Introduction}

The topic of pattern formation, or the spontaneous emergence of ordered
structures, is nowadays an active field of research in many areas of nonlinear
science \cite{Cross93}. Pattern formation is commonly observed in large aspect
ratio nonlinear systems which are driven far from the equilibrium state by an
external input. Transverse modes may become unstable when the amplitude of the
external input reaches a critical threshold value, large enough to overcome
the losses produced by dissipative processes in the system, and a symmetry
breaking transition developes, carrying the system from an initially
homogeneous to an inhomogeneous state, usually with spatial periodicity.

Parametrically driven systems offer many examples of spontaneous pattern
formation. For example, parametric excitation of surface waves by a vertical
shake (Faraday instability) in fluids \cite{Miles84} and granular layers
\cite{Melo94}, spin waves in ferrites and ferromagnets and Langmuir waves in
plasmas parametrically driven by a microwave field \cite{Lvov94}, or the
optical parametric oscillator \cite{Oppo94,deValcarcel96} have been studied.

In nonlinear acoustics, a phenomenon which belongs to the class of the
previous examples is the parametric sound amplification. It consists in the
resonant interaction of a triad of sound waves with frequencies $\omega
_{0},\omega_{1}$ and $\omega_{2}$, for which the following energy and momentum
conservation conditions are fulfilled:
\begin{align}
\omega_{0}  &  =\omega_{1}+\omega_{2},\nonumber\\
\vec{k}_{0}  &  =\vec{k}_{1}+\vec{k}_{2}+\Delta\vec{k}, \label{conservation}%
\end{align}
where $\Delta\vec{k}$ is a small phase mismatch. The process is initiated by
an input pumping wave of frequency $\omega_{0}$ which, due to the coupling to
the nonlinear medium, generates a pair of waves with frequencies $\omega_{1}$
and $\omega_{2}$. When the wave interaction occurs in a resonator, a threshold
value for the input amplitude is required, and the process is called
parametric sound generation. In acoustics, this process has been described
before by several authors under different conditions, either theoretical and
experimentally. In \cite{Korpel65,Adler70,Yen75} the one dimensional case
(colinearly propagating waves) is considered. In \cite{Ostrovsky76} the
problem of interaction between concrete resonator modes, with a given
transverse structure, is studied. In both cases, small aspect ratio resonators
containing liquid and gas respectively are considered. More recently,
parametric interaction in a large aspect ratio resonator filled with
superfluid He$^{4}$ has been investigated \cite{Rinberg96}.

It is well known that optical and acoustical waves share many common
phenomena, an analogy which sometimes can be extended to the nonlinear regime
\cite{Bunkin86}. In particular, the phenomenon of parametric sound generation
is analogous to the optical parametric oscillation in nonlinear optics.
However, an important difference between acoustics and optics is the absence
of dispersion in the former. In a nondispersive medium, all the harmonics of
each initial monochromatic wave propagate synchronously. As a consequence, the
spectrum broadens during propagation and the energy is continuously pumped
into the higher harmonics, leading to waveform distortion and eventually to
the formation of shock fronts. On the contrary, dispersion allows that only
few modes, those satisfying given synchronism conditions, participate
effectively in the interaction process.

In acoustics, the presence of higher harmonics can be avoided by different
means. One method is based in the introduction of some dispersion mechanism.
In finite geometries, such as waveguides \cite{Hamilton87} or resonators
\cite{Ostrovsky78}, the dispersion is introduced by the lateral boundaries.
Different resonance modes, propagating at different angles, propagate with
different effective\ phase velocities. Other proposed methods are, for
example, the inclusion of media with selective absorption, in which selected
spectral components experience strong losses and may be removed from the wave
field \cite{Zarembo74}, or resonators where the end walls present a
frequency-dependent complex impedance \cite{Yen75}. In this case, the
resonance modes of the resonator are not integrally related, and by proper
adjustment of the resonator parameters one can get that only few modes, those
lying close enough to a cavity resonance, reach a signifficant amplitude. In
any of these cases, a spectral approach to the problem, in terms of few
interacting modes, is justified.

Therefore the selective effect of the resonator allows to reduce the study of
parametric sound generation to the interaction of three field modes,
correspondig to the driving (fundamental) and subharmonic frequencies, and to
describe this interaction through a small set of nonlinear coupled
differential equations. In the present work we concentrate on the particular
degenerate case of subharmonic generation, where $\omega_{1}=\omega_{2}$ and
consequently $\omega_{0}=2\omega_{1},$ being $\omega_{0}$ the fundamental and
$\omega_{1}$ generated subharmonic, both quasi-resonant with a corresponding
resonance mode. This degenerate case has been considered in previous
experimental studies \cite{Yen75,Ostrovsky78}, in the case of small aspect
ratio cavities where transverse spatial evolution is absent.

The aim of the paper is twofold. On one side, a rigurous derivation of the
dynamical model describing the parametric interaction of acoustic waves in a
large aspect ratio cavity is presented. The derived model is isomorphous to
the system of equations describing parametric oscillation in an optical
resonator, and consequently their solutions are known. On the other side, the
model predicts the existence of pattern forming or modulational instabilities,
when the subharmonic field is negatively detuned with respect to the cavity.
The second aim of the paper is therefore to determine the conditions under
which pattern formation of sound could be observed in an acoustic system.
Numerical integrations under realistic acoustical parameters confirm the
predicted results.

\section{Derivation of the model}

\subsection{Three wave interaction in an acoustic resonator}

The physical system we consider in this paper is an acoustic cavity
(resonator) composed by two parallell solid walls, with thicknesses $D$ and
$H$ separated a distance $L$, containing a fluid medium inside, as described
in Fig. 1. The different media are acoustically characterized by its density
$\rho$ and the propagation velocity of the sound wave, $c$. One of the walls
vibrates at a frequency close to one of the normal modes of the cavity.

The resonance modes $f$ (eigenfrequencies) of such resonator can be calculated
by using the equation \cite{Yen75}%
\begin{equation}
\mathcal{R}\left(  \tan\frac{f}{f_{D}}\pi+\tan\frac{f}{f_{H}}\pi\right)
+\left(  1-\mathcal{R}^{2}\tan\frac{f}{f_{D}}\pi~\tan\frac{f}{f_{H}}%
\pi\right)  \tan\frac{f}{f_{L}}\pi=0, \label{modes2}%
\end{equation}
where $\mathcal{R=}\rho_{w}c_{w}/\rho c$ is the ratio of wall to medium
acoustic impedances, $f_{D}=c_{w}/2D,$ $f_{H}=c_{w}/2H,$\ and $f_{L}=c/2L$ are
the fundamental resonance frequencies of each individual region, respectively.
From the numerical solutions of Eq. (\ref{modes2}) results a non-equidistant
spectrum, the position of the different modes being determined by the
properties and dimensions of the different elements. Note that the particular
case corresponding to an equidistant spectrum (constant free spectral range)
is obtained as a limit case when we impose infinite reflectance at the walls
($\mathcal{R}\rightarrow\infty$) with negligible thickness. In this case Eq.
(\ref{modes2}) reduces to $\tan kL=0$, and the modes obey the Fabry-Perot
condition $k=n\pi/L$. In such a perfect resonator, any harmonic of a resonant
driving wave is also resonant with a higher--order cavity mode, and the energy
flow into these modes leads to wave distortion and invalidates a modal
description of the problem in terms of the interaction among few waves.

In a loosy resonator, however, one can get the second harmonic of the driving
wave to be more detuned than subharmonics with respect to a cavity resonance,
thus reducing the effectivity of the cascade process into the higher
harmonics. This effect is enhanced in the case of viscous media, in which the
higher frequencies experience stronger losses (absorption). Furthermore, it is
possible to get subharmonic generation slightly detuned from a cavity
resonance, which is a necessary condition for the developement of spatial
instabilities, as will be discussed in the following sections. These two facts
justify the description of high intensity acoustic waves in a resonator in
terms of the interaction among few frequency components. Several experimental
results demonstrate this fact \cite{Korpel65,Adler70,Yen75}, and support the
validity of this assumption in the theoretical approach.

The main novelty of this work with respect to previous studies, is to consider
the diffraction of the waves inside the cavity. Diffraction can play an
important role when the cavity has a large Fresnel number, defined as
$F=a^{2}/\lambda L,$ where $a$ is the characteristic transverse size of the
cavity (for example, $a^{2}$ is the area of a plane radiator), $\lambda$ is
the wavelength and $L$ is the length of the cavity in the direction of
propagation, considered the longitudinal axis of the cavity. Sometimes the
case of large $F$ is called the large aspect ratio limit. All these
assumptions will be taken into account in the derivation of the model in the
next section.

\subsection{Hydrodynamic equations for sound waves}

As a starting point of the analysis, we consider the basic hydrodinamic
equations describing the propagation of sound waves in liquids and gases,
namely the continuity (mass conservation) equation,%
\begin{equation}
\frac{\partial\rho}{\partial t}+\mathbf{\nabla}(\rho\mathbf{u})=0,
\label{continuity}%
\end{equation}
and the Euler (momentum conservation) equation,%
\begin{equation}
\rho\left(  \frac{\partial}{\partial t}+\mathbf{u\nabla}\right)
\mathbf{u}=-\mathbf{\nabla}p+\mu\nabla^{2}\mathbf{u}+(\mu_{B}+\frac{\mu}%
{3})\mathbf{\nabla(\nabla u),} \label{euler}%
\end{equation}
where $\rho$ is the density of the medium, $\mathbf{u}$ is the fluid particle
velocity, $p$ is the thermodinamic pressure, and $\mu$ and $\mu_{B}$ represent
shear and bulk viscosities, respectively. Equations (\ref{continuity}) and
(\ref{euler}) must be complemented by the equation of state $p=p(\rho)$. If
the losses due to viscosity are small (due just to heat conduction) the
process can be assumed to be adiabatic. Then the pressure in the state
equation can be expanded around the equilibrium and the equation of state
takes the form%
\begin{align}
p  &  =p_{0}+\left(  \frac{\partial p}{\partial\rho}\right)  _{s}\rho^{\prime
}+\frac{1}{2}\left(  \frac{\partial^{2}p}{\partial\rho^{2}}\right)  _{s}%
^{2}\rho^{\prime~2}+...=\nonumber\\
&  =p_{0}+c_{0}^{2}\rho^{\prime}+\frac{1}{2}\Gamma\rho^{\prime~2}+...,
\label{state}%
\end{align}
where $\rho^{\prime}=\rho-\rho_{0}$, being $\rho_{0}$ the equilibrium value of
the density, $c_{0}=\sqrt{(\partial p/\partial\rho)_{s}}$ is the (low
amplitude) sound velocity and
\[
\Gamma=\left(  \frac{\partial^{2}p}{\partial\rho^{2}}\right)  _{s}=\frac
{c_{0}^{2}}{\rho_{0}}\frac{B}{A},
\]
where $B/A$ is commonly used in acoustics as the nonlinearity parameter, and
has been measured in different media \cite{Hamiltonbook}. The subscript $s$
denotes the adiabatic character of the process, and the ellipsis in Eq.
(\ref{state}) the nonlinearities higher than quadratic, which are neglected.

Substitution of Eq. (\ref{state}) in (\ref{euler}) leads to%
\begin{equation}
\rho\left(  \frac{\partial}{\partial t}+\mathbf{u\nabla}\right)
\mathbf{u}=-c_{0}^{2}\mathbf{\nabla}\rho^{\prime}-\frac{1}{2}\Gamma
\mathbf{\nabla}\rho^{\prime~2}\mathbf{+}\mu\nabla^{2}\mathbf{u}+(\mu_{B}%
+\frac{\mu}{3})\mathbf{\nabla(\nabla u).} \label{euler2}%
\end{equation}
which together with Eq. (\ref{continuity}) are a two-variable model. It is
convenient to write Eqs. (\ref{continuity}) and (\ref{euler2}) in
nondimensional form, adopting the following normalizations:%
\begin{equation}
\mathbf{v}\equiv\frac{\mathbf{u}}{V},~\bar{\rho}\equiv\frac{\rho}{\rho_{0}},
\label{norm1}%
\end{equation}
where $V$ is a reference velocity, small compared with $c_{0}$. Also, time and
space are defined as
\begin{equation}
\bar{t}=\omega t,~\mathbf{\bar{x}}=k\mathbf{x}. \label{norm2}%
\end{equation}
where $\omega$ and $k$ are the angular frequency and wave number of a
reference wave, and obey $\omega=kc_{0}.\ $With this normalization Eqs.
(\ref{continuity}) and (\ref{euler2}) have the form%
\begin{equation}
\frac{\partial\bar{\rho}}{\partial\bar{t}}+M\rho~\mathbf{\bar{\nabla}%
v}+M\mathbf{v~\bar{\nabla}}\bar{\rho}=0, \label{continuity2}%
\end{equation}%
\begin{equation}
M\bar{\rho}\frac{\partial\mathbf{v}}{\partial\bar{t}}+M^{2}\bar{\rho
}\mathbf{v\bar{\nabla}~v=-\bar{\nabla}}\bar{\rho}-\frac{1}{2}\bar{\Gamma
}\mathbf{\bar{\nabla}}\left(  \bar{\rho}-1\right)  ^{2}+\bar{\mu}~M^{2}%
\bar{\nabla}^{2}\mathbf{v.} \label{euler3}%
\end{equation}
where the losses $\bar{\mu}$, the nonlinearity $\bar{\Gamma}$ and the acoustic
Mach number $M$, are parameters defined as%
\begin{align}
\bar{\mu}  &  =\frac{k}{\rho_{0}V}(\mu_{B}+\frac{4}{3}\mu),\\
\bar{\Gamma}  &  =\frac{\rho_{0}}{c_{0}^{2}}\Gamma\equiv\frac{B}{A},\\
M  &  =\frac{V}{c_{0}}.
\end{align}

In Eq. (\ref{euler3}) we have used the identity $\mathbf{\nabla(\nabla
v)=}\nabla^{2}\mathbf{v+\nabla\times\nabla\times v}$, where the second
(vorticity) term has been neglected, since its magnitude decays exponentially
away from the boundaries \cite{Hamiltonbook}.

\subsection{Perturbative expansion in the small Mach number limit}

Under usual conditions, the acoustic Mach number take small values
($M<10^{-3}$), which allows to treat Eqs. (\ref{continuity}) and
(\ref{euler2}) by perturbative techniques. Thus we consider a smallness
parameter $\varepsilon$ as the Mach number, and express the parameters and
variables in terms of it.

Let us assume that in the dispersive resonator, the changes in the shape of
the wave as a consequence of dissipation and nonlinearity, both along the
direction of propagation and transverse to it, are small. Also, we take into
account that the changes along the transverse direction to the propagation,
due to diffraction, take place faster than along this propagation direction
\cite{Bakhvalov}. These assumptions allow to consider the problem in terms of
fast and slow scales. A choice of scales accounting for these changes is
\begin{subequations}
\label{scales1}%
\begin{align}
\bar{t}  &  =T+\varepsilon\tau,\\
\left(  \bar{z},\bar{x},\bar{y}\right)   &  =\left(  z,\sqrt{\varepsilon
}x,\sqrt{\varepsilon}y\right)  ,
\end{align}
and expand the state variables $\bar{\rho}$ and $\mathbf{v=(}v_{x},v_{y}%
,v_{z}\mathbf{)}$ as
\end{subequations}
\begin{subequations}
\label{scales2}%
\begin{align}
\bar{\rho}  &  =1+\varepsilon\rho_{1}+\varepsilon^{2}\rho_{2},\\
v_{z}  &  =v_{1z}+\varepsilon v_{2z},\\
v_{x}  &  =\sqrt{\varepsilon}v_{1x}+\varepsilon\sqrt{\varepsilon}v_{2x},\\
v_{y}  &  =\sqrt{\varepsilon}v_{1y}+\varepsilon\sqrt{\varepsilon}v_{2y}.
\end{align}
where the order of the transverse components of the velocity is determined by
the (slow) divergence of the beam \cite{Bakhvalov}. Note that, since at
equilibrium fluid is at rest and, on other hand, $v$ is nondimensionalized by
$V$ (a reference velocity smaller than $c_{0}$), the variation of $v$ is order
$\mathcal{O}(1)$. Substituting these expressions into Eqs. (\ref{continuity2})
and (\ref{euler3}), we obtain equations at different orders which can be
recursively solved. The leading order $\mathcal{O}(\varepsilon) $ reads
\end{subequations}
\begin{subequations}
\label{order0}%
\begin{align}
\frac{\partial v_{1z}}{\partial T}  &  =-\frac{\partial\rho_{1}}{\partial
z},\label{order0a}\\
\frac{\partial\rho_{1}}{\partial T}  &  =-\frac{\partial v_{1z}}{\partial z}.
\label{order0b}%
\end{align}
which leads to linear wave equations for the particle velocity and density. In
general, the frequencies of the waves are not resonant with a cavity
eigenmode, and the waves at any frequency are detuned, the detuning parameter
being defined as
\end{subequations}
\begin{equation}
\delta_{i}=\omega_{i}^{c}-\omega_{i}, \label{detuning}%
\end{equation}
where $\omega_{i}$ is the frequency of the field and $\omega_{i}^{c}$ is
frequency of the cavity eigenmode closest to $\omega_{i}$. In this case, the
solution of the order $\mathcal{O}(\varepsilon)$ takes the general form of a
superposition of standing waves%
\begin{equation}
v_{1z}=\underset{n=0}{\overset{2}{%
{\displaystyle\sum}
}}A_{n}(x,y,\tau)\sin[\omega_{n}T-\left(  \phi_{n}(\tau)-\delta_{i}T\right)
]\sin\left(  k_{n}z\right)  . \label{standing}%
\end{equation}
where $\omega_{n}=k_{n}$. In Eq. (\ref{standing}) we have considered the
frequency--selective effect of the walls discussed in the introduction, and
assume that only three modes, those with frequencies obeying $\omega
_{1}+\omega_{2}=\omega_{0}$, can reach signifficant amplitudes. Also, from
Eqs. (\ref{order0}) we obtain that%
\begin{equation}
\rho_{1}=\underset{n=0}{\overset{2}{%
{\displaystyle\sum}
}}A_{n}(x,y,\tau)\cos[\omega_{n}T-\left(  \phi_{n}(\tau)-\delta_{i}T\right)
]\cos\left(  k_{n}z\right)  . \label{standing2}%
\end{equation}

At order $\mathcal{O}(\varepsilon^{3/2})$, the equations for the fast
evolution of the transverse velocity components are obtained,
\begin{subequations}
\label{order32}%
\begin{align}
\frac{\partial v_{1x}}{\partial T}  &  =-\frac{\partial\rho_{1}}{\partial
x};\label{order32a}\\
\frac{\partial v_{1y}}{\partial T}  &  =-\frac{\partial\rho_{1}}{\partial y}.
\label{order32b}%
\end{align}

At order $\mathcal{O}(\varepsilon^{2})$ we get
\end{subequations}
\begin{subequations}
\label{order2}%
\begin{align}
\rho_{1}\frac{\partial v_{1z}}{\partial T}+\frac{\partial v_{2z}}{\partial
T}+\frac{\partial v_{1z}}{\partial\tau}+\frac{1}{2}\frac{\partial^{2}%
v_{1z}^{2}}{\partial z^{2}}+\frac{\bar{\Gamma}}{2}\frac{\partial\rho_{1}^{2}%
}{\partial z}-\bar{\mu}\frac{\partial^{2}v_{1z}}{\partial z^{2}}%
+\frac{\partial\rho_{2}}{\partial z}  &  =0,\label{order2a}\\
\frac{\partial\rho_{2}}{\partial T}+\frac{\partial\rho_{1}}{\partial\tau}%
+\rho_{1}\frac{\partial v_{1z}}{\partial z}+\frac{\partial v_{2z}}{\partial
z}+v_{1z}\frac{\partial\rho_{1}}{\partial z}+\frac{\partial v_{1x}}{\partial
x}+\frac{\partial v_{1y}}{\partial y}  &  =0. \label{order2b}%
\end{align}

Finally Eqs. (\ref{order2}), making use of the relations in Eqs.
(\ref{order32}), can be reduced to a single wave equation for the density
\end{subequations}
\begin{equation}
\frac{\partial^{2}\rho_{2}}{\partial T^{2}}-\frac{\partial^{2}\rho_{2}%
}{\partial z^{2}}=2\frac{\partial^{2}v_{1z}}{\partial z\partial\tau}-\bar{\mu
}\frac{\partial^{3}v_{1z}}{\partial z^{3}}+\frac{\bar{\Gamma}}{2}%
\frac{\partial^{2}\rho_{1}^{2}}{\partial z^{2}}+\frac{\partial^{2}v_{1z}^{2}%
}{\partial z^{2}}+\frac{\partial^{2}\rho_{1}}{\partial x^{2}}+\frac
{\partial^{2}\rho_{1}}{\partial y^{2}}. \label{order22}%
\end{equation}
where the smaller order solutions appear at the right--hand side as source terms.

A closed set of equations for the slowly varying envelopes of mode amplitudes
$A_{n}$ and phases $\phi_{n}$ can be obtained by substituting Eqs.
(\ref{standing}) and (\ref{standing2}) in Eq. (\ref{order22}), and imposing
the absence of secular terms in the resulting equation, i.e. neglecting the
source contributions that contain the same frequency components as the natural
frequency of the left--hand side part in Eq. (\ref{order22}). Otherwise,
second order solutions would grow linearly, violating the smallness condition
$\varepsilon\rho_{2}<<\rho_{1}$ assumed in the perturbation expansion. After
some algebra the secular terms reduce to the following real system for the
amplitudes and phases
\begin{subequations}
\label{eqmodes}%
\begin{align}
\frac{\partial A_{i}}{\partial\tau}+\sigma s_{i}\omega_{i}A_{j}A_{k}\sin
\phi+\frac{1}{2}\bar{\mu}\omega_{i}^{2}A_{i}  &  =0,\\
A_{i}\frac{\partial\phi_{i}}{\partial\tau}-\sigma\omega_{i}A_{j}A_{k}\cos
\phi-\frac{1}{2\omega_{i}}\nabla_{\bot}^{2}A_{i}-\delta_{i}A_{i}  &  =0,
\end{align}
where $(i,j,k)=(0,1,2)$, and the other two equations are obtained by cyclic
permutations. The sign operator $s_{i}=+1$ for $i=1,2$ and $-1$ for $i=0.$ A
global phase is defined as $\phi\equiv\phi_{1}+\phi_{2}-\phi_{0}$,
$\nabla_{\bot}^{2}$ stems for the Laplacian operator acting on the transverse
space $\mathbf{r}_{\bot}=(x,y)$, and $\sigma$ is a coupling parameter defined
as
\end{subequations}
\begin{equation}
\sigma=\frac{1}{4}\left(  1+\frac{B}{2A}\right)  . \label{coupling}%
\end{equation}

Note that in Eqs. (\ref{eqmodes}) the terms proportional to $\bar{\mu}$
accounts only for viscous losses. There are in fact other loss mechanisms,
mainly related with finite reflectance of the walls or diffraction losses
through the open sides of the resonator. When the losses are sufficiently
small, one can generalize Eqs. (\ref{eqmodes}) and consider an effective
(phenomenological) loss parameter $\gamma_{i}$ for each mode, just replacing
$\frac{\bar{\mu}}{2}\omega_{n}^{2}A_{n}$ by $\gamma_{n}A_{n}$. The value of
these coefficents can be obtained experimentally for a particular resonator by
small amplitude measurements of the decay rate of a given mode, since under
this condition (neglecting nonlinearity) the amplitudes obey $\partial
A_{n}/\partial\tau=-\gamma_{n}A_{n}.$

Finally, for a dissipative resonator, an external source must be provided in
order to compensate the losses. Consider that a plane wave of amplitude $E$
and frequency $\omega_{0}^{c}$ is injected in the resonator. In each
roundtrip, the amplitude of the standing wave will increase by $2E$, and then%
\begin{equation}
\frac{\Delta A_{0}}{\Delta t}=\frac{2E}{2L/c_{0}}=\frac{c_{0}}{L}E
\label{pump}%
\end{equation}
where $L$ is the length of the resonator and $c_{0}/L$ corresponds to the time
taken for a wave to travel across it. By changing to the nondimensional
notation in slow time scale, and assuming small amplitude changes during a
roundtrip, one can consider the differential limit of Eq. (\ref{pump}) and
incorporate it to the evolution equation of $A_{0}$ as a driving term.

A particular case corresponds to degenerate interaction, where $\omega
_{1}=\omega_{2}$. This process describes the subharmonic parametric
generation. In this situation, the pump and subharmonic waves obey the
relation $\omega_{0}=2\omega_{1}$. It is worth to express the resulting system
of equations in complex form, by defining the complex amplitudes as
$B_{n}(\mathbf{r}_{\bot},\tau)=A_{n}(\mathbf{r}_{\bot},\tau)\exp\left[
i\phi_{n}(\tau)\right]  .$ At this point we also return to dimensional
(physical) variables. Since amplitudes $A_{n}$ correspond, e.g. to
dimensionless densities, and pressure is related to density by Eq.
(\ref{state}), then, in the degenerate limit we obtain the following coupled
equations for the evolutions of pressure
\begin{subequations}
\label{model1}%
\begin{align}
\frac{\partial p_{0}}{\partial t}  &  =-(\gamma_{0}+i\delta_{0})p_{0}%
-i\frac{\sigma\omega_{0}}{\rho_{0}c_{0}^{2}}p_{1}^{2}+i\frac{c_{0}^{2}%
}{2\omega_{0}}\nabla_{\bot}^{2}p_{0}+\frac{c_{0}}{L}p_{in},\\
\frac{\partial p_{1}}{\partial t}  &  =-(\gamma_{1}+i\delta_{1})p_{1}%
-i\frac{\sigma\omega_{1}}{\rho_{0}c_{0}^{2}}p_{1}^{\ast}p_{0}+i\frac{c_{0}%
^{2}}{2\omega_{1}}\nabla_{\bot}^{2}p_{1}.
\end{align}
together with their complex conjugate. In Eqs. (\ref{model1}), $p_{i}$
corresponds to deviations with respect to equilibrium pressure values.

Equations (\ref{model1}) can be further simplified by adopting the following
normalizations
\end{subequations}
\begin{subequations}
\label{normalizations}%
\begin{align}
p_{0}  &  =i\frac{2\rho_{0}c_{0}^{2}\gamma_{1}}{\sigma\omega_{0}}%
\mathcal{P}_{0},\\
p_{1}  &  =\frac{\rho_{0}c_{0}^{2}\sqrt{2\gamma_{0}\gamma_{1}}}{\sigma
\omega_{0}}\mathcal{P}_{1},\\
p_{in}  &  =i\frac{2L\rho_{0}c_{0}\gamma_{0}\gamma_{1}}{\sigma\omega_{0}%
}\mathcal{E}%
\end{align}
and introducing the dimensionless detuning parameter $\Delta_{n}=\delta
_{n}/\gamma_{n}.$ The final form of the model reads
\end{subequations}
\begin{subequations}
\label{model2}%
\begin{align}
\frac{1}{\gamma_{0}}\frac{\partial\mathcal{P}_{0}}{\partial t}  &
=-(1+i\Delta_{0})\mathcal{P}_{0}-\mathcal{P}_{1}^{2}+ia_{0}\nabla_{\bot}%
^{2}\mathcal{P}_{0}+\mathcal{E},\\
\frac{1}{\gamma_{1}}\frac{\partial\mathcal{P}_{1}}{\partial t}  &
=-(1+i\Delta_{1})\mathcal{P}_{1}+\mathcal{P}_{1}^{\ast}\mathcal{P}_{0}%
+ia_{1}\nabla_{\bot}^{2}\mathcal{P}_{1}.
\end{align}
where $a_{n}=c_{0}^{2}/2\omega_{n}\gamma_{n}$ are the diffraction
coefficients. This form of the equations is relevant for our purposes, since
their solutions and stability have been discussed in the context of nonlinear
optics, after the model given by Eqs. (\ref{model2}) has been derived for the
degenerate optical parametric oscillator. A detailed analysis of the
spatio--temporal dynamics of Eqs. (\ref{model2}) has been carried out during
the last decade (starting with the seminal work \cite{Oppo94}), and a recent
overview can be found in \cite{springerbook}. In the following sections, we
review the basic results regarding their homogeneous solutions and their
stability, and we study numerically the spatio--temporal dynamics under
conditions corresponding to a real acoustical resonator. Note that the
connection between the generic model (\ref{model2}) and the particular
acoustic problem is given by the normalizations performed in Eqs.
(\ref{normalizations}).

\section{Modulational instabilities of homogeneous solutions}

Two stationary states are solution of Eqs. (\ref{model2}): the simplest,
trivial solution,
\end{subequations}
\begin{equation}
\overline{\mathcal{P}}_{0}=\frac{\mathcal{E}}{\left(  1+i\Delta_{0}\right)
},\,\overline{\mathcal{P}}_{1}=0, \label{trivial}%
\end{equation}
characterized by a null value of the subharmonic field inside the resonator,
and the nontrivial solution
\begin{subequations}
\label{notrivial}%
\begin{align}
\left\vert \overline{\mathcal{P}}_{0}\right\vert ^{2}  &  =1+\Delta_{1}^{2},\\
\left\vert \overline{\mathcal{P}}_{1}\right\vert ^{2}  &  =-1+\Delta_{0}%
\Delta_{1}\pm\sqrt{\left\vert \mathcal{E}\right\vert ^{2}-\left(  \Delta
_{0}+\Delta_{1}\right)  ^{2}},
\end{align}
in which both the pump and the subharmonic fields have a nonzero amplitude,
and exists above a given (threshold) pump value $\mathcal{E=E}_{th}$. At this
value, given by
\end{subequations}
\begin{equation}
\left\vert \mathcal{E}_{th}\right\vert =\sqrt{\left(  1+\Delta_{0}^{2}\right)
\left(  1+\Delta_{1}^{2}\right)  }, \label{1erumbral}%
\end{equation}
the trivial solution loose its stability and bifurcates into the nontrivial
one. The emergence of this finite amplitude solution corresponds to the
process of subharmonic generation. Note that the fundamental amplitude above
the threshold is independent of the value of the injected pump, which means
that all the energy is transferred to the subharmonic wave.

These results have been confirmed experimentally for an acoustical resonator
in \cite{Ostrovsky76}. The character of the bifurcation depends on the
detuning values. As demonstrated in \cite{Ostrovsky76}, and also in the
optical context, \cite{Lugiato88} the bifurcation is supercritical when
$\Delta_{0}\Delta_{1}<1$, and subcritical when $\Delta_{0}\Delta_{1}>1$. In
the latter case, both trivial and finite amplitude solutions can coexist for
given sets of the parameters, which results in a regime of bistability between
different solutions.

In order to study the stability of the trivial solution (\ref{trivial})
against space-dependent perturbations, consider a deviation of this state,
given by $A_{j}\left(  \mathbf{r}_{\mathbf{\perp}},t\right)  =\bar{A}%
_{j}+\delta A_{j}\left(  \mathbf{r}_{\mathbf{\perp}},t\right)  .$ Assuming
that the deviations are small with respect to the stationary values, one can
substitute the perturbed solution in Eqs.(\ref{model2}) and linearize the
resulting system in the perturbations $\delta A_{j}.$ The generic solutions of
the linear system are of the form
\begin{equation}
\left(  \delta A_{j},\delta A_{j}^{\ast}\right)  \propto e^{\lambda\left(
k_{\perp}\right)  t}e^{i\mathbf{k}_{\perp}\cdot\mathbf{r}_{\perp}},
\label{perturb}%
\end{equation}
where $\lambda(\mathbf{k}_{\mathbf{\perp}})$ represents the growth rate of the
perturbations, and $\mathbf{k}_{\mathbf{\perp}}$ is the transverse component
of the wavevector, which in a two-dimensional geometry obeys the relation
$\left\vert \mathbf{k}_{\mathbf{\perp}}\right\vert ^{2}=k_{x}^{2}+k_{y}^{2}.$
The growth rates, which depend on the wavenumber of the perturbations, are
obtained as the eigenvalues of the linear system. This analysis has been
performed before \cite{Oppo94}, and we present here the main conclusions,
omitting details.

The eigenvalue (and consequently the instabilities) presents a different
character depending on the sign of the subharmonic detuning. If $\Delta_{1}%
>0$, which corresponds to a subharmonic frequency smaller than that of the
closest cavity mode, the eigenvalue shows a maximum at $k_{\perp}=0,$ the
emerging solution being homogeneous in transverse space, with amplitude given
by Eq.(\ref{notrivial}). On the contrary, in the opposite case $\Delta_{1}<0$,
which corresponds to field frequencies larger than the nearest cavity mode,
the maximum of the eigenvalue occurs for perturbations with transverse
wavenumber
\begin{equation}
k_{\perp}=\sqrt{-\frac{\Delta_{1}}{a_{1}}}. \label{kpattern}%
\end{equation}

The emerging solution in this case is of the form Eq.(\ref{perturb}), which
represents a plane wave tilted with respect to the cavity axis. This solution
presents spatial variations in the transverse plane, and consequently pattern
formation is expected to occur.

Since $k_{\perp}$ is the modulus of the wavevector, the linear stability
analysis in two dimensions predicts that a continuum of modes within a
circular annulus (centered on a critical circle at $\left|  \mathbf{k}_{\perp
}\right|  =k_{\perp}$ in $\left(  k_{x},k_{y}\right)  $ space) grows
simultaneously as the pump increases above a critical value. This double
infinite degeneracy of spatial modes (degenerate along a radial line from the
origin and orientational degeneracy) allows, in principle, arbitrary
structures in two dimensions.

The threshold for pattern formation follows also from the eigenvalue, and is
given by
\begin{equation}
\mathcal{E}_{p}=\sqrt{1+\Delta_{0}^{2}}. \label{threshold3}%
\end{equation}

The predictions of the stability analysis correspond to the linear stage of
the evolution, where the subharmonic field amplitude is small enough to be
considered a perturbation of the trivial state. The analytical study of the
further evolution would require a nonlinear stability analysis, not given
here. Instead, in the next section we perform the numerical integration of
Eqs.(\ref{model2}), where predictions of the acoustic subharmonic\ field in
the linear and nonlinear regime are given.

\section{Numerical results in the acoustical case}

The analytical predictions of the linear stability analysis have been
numerically confirmed for Eqs. (\ref{model2}) in previous studies, in the
context of nonlinear optics. In this section we demonstrate the adequacy of
these resuls for the acoustical case. For this aim, we first evaluate the
different parameters appearing in Eqs. (\ref{model1}) for a concrete case.

Consider a resonator composed by two identical walls of thickness $D=H=0.5$ cm
made of a lead zirconate titanate (PZT) piezoelectric material ($c_{t}=4400$
m/s$,\rho_{t}=7700$ kg/m$^{\text{3}}$), containing water ($c_{m}=1480$
m/s$,\rho_{t}=1000$ kg/m$^{\text{3}}$). For this case $\mathcal{R}=22.89.$ The
length of the medium $L$ can be varied in order to modify detunings. If the
resonator is driven at a frequency $f_{0}=4$ MHz, then subharmonic generation
is expected to occur at $f_{1}=2$ MHz. The corresponding detunings have been
numerically evaluated from Eq. (\ref{modes2}). For a cavity length $L=3$ cm,
the pump is almost resonant with a cavity mode, $\delta f_{0}=f_{0}^{c}%
-f_{0}\approx0$ KHz, and the subharmonic is detuned by $\delta f_{1}=f_{1}%
^{c}-f_{1}\approx-1.6$ KHz. Furthermore, under these conditions the second
harmonic at $f_{2}=8$ MHz is highly detuned, by $\delta f_{2}=f_{2}^{c}%
-f_{2}\approx-3.7$ KHz, and therefore it will reach a small amplitude.

The loss coefficients, as stated before, can be obtained for small amplitude
measurements of the decay rate of each mode in the resonator. In particular, a
measurement of the quality factor for the different cavity modes, defined as
$Q_{i}=\omega_{i}/2\gamma_{i}$ where performed in \cite{Yen75} for a similar
interferometer. For the frequencies of interest, the measured quality factors
take values of the order of $10^{3}-10^{4}.$ From this results we can conclude
that reasonable values for the decay rates are $\gamma_{0}=\gamma_{1}%
=5~10^{3}$ rad/s, which allows to evaluate the rest of parameters in the
model. The normalized detuning parameters corresponding to this case are
$\Delta_{0}=0$ and $\Delta_{1}=-1.6,$ and diffraction coefficients result
$a_{0}=8.7~10^{-6}$ and $a_{1}=2a_{0}\,\ $Finally, the nonlinearity parameter
of water at 20%
${{}^o}$%
C is $B/A=5$, which substituted in Eq. (\ref{coupling}) gives $\sigma=0.875. $

The theory of the previous section predicts that, when the threshold value Eq.
(\ref{threshold3}) is achieved, a periodic pattern with a characteristic scale
given by Eq. (\ref{kpattern}) develops. For the above conditions, the
normalized threshold value is $\mathcal{E}_{p}=1$, and the corresponding input
pressure at the driving wall is obtained from the last of Eqs.
(\ref{normalizations}), and results $p_{in}\approx0.1~$MPa. Also, the
wavelength of the pattern is obtained from $\lambda_{\bot}=2\pi/k_{\bot
}\approx1.5$ cm. Then, in order to observe the pattern the transverse section
must contain several wavelengths, which in turn which implies a transverse
size of 10 cm or more. All these values can be considered as realistic.

In order to check the analytical predictions, we integrated numerically
Eqs.(\ref{model2}) by using the split-step technique on a spatial grid of
dimensions 128$\times$128 \cite{deValcarcel96}. The local terms, either linear
(pump, losses and detuning) and nonlinear, are calculated in the space domain,
while nonlocal terms (diffractions) are evaluated in the spatial wavevector
(spectral) domain. A Fast Fourier Transform (FFT) is used to shift from
spatial to spectral domains in every time step. Periodic boundary conditions
are used.

As a initial contition, a noisy spatial distribution is considered, and the
parameters are those discussed above, for which a pattern forming instability
is predicted. Figure 2 shows the result of the numerical integration. In Figs.
2(a) and (b) several snapshots of the evolution at different times are shown,
which eventually result in a final stable one-dimensional pattern in the form
of stripes, shown in Fig. 2(c).

Numerical simulations for different detunings have been performed. A
systematic study shows that in most of the cases the system developes stripped
patterns with arbitrary orientations. However, in some cases a pattern with
squared symmetry result as the final stable state. An example of evolution
leading to squared patterns is shown in Fig. (3), obtained for $\Delta_{1}%
=-4$, $\Delta_{0}=-8$ and $\mathcal{E}=1.2.$

\section{Conclusions}

The pattern formation properties of an acoustical resonator where subharmonic
generation takes place are discussed from the theoretical point of view. A
model allowing for diffraction of the fields (large-aspect ratio limit) is
derived by means of the multiple scale expansions technique. The obtained
model, which is isomorphous to that obtained for the optical parametric
oscillator, is analized in detail considering the distinctive peculiarities of
the acoustical system. A typical acoustical configuration is considered, and
the predictions of the theory are confirmed by numerical integration under
realistic conditions. Numerics show that transverse patterns in the form of
one-dimensional stripes are usually obtained as the final stable state,
although the system can support also patterns with more complex structures,
such as squares.

We finally note that there has been several experimental attempts to verify
the pattern formation scenario predicted by Eqs. (\ref{model2}) in a nonlinear
optical resonator \cite{Ducci01,Vaupel99}. Although some transverse patterns
with different latticelike structures have been observed, the use of other
cavities different than planar (confocal and concentric) was required. Owing
to the complete analogy between the model derived in this paper, and the model
describing optical parametric oscillation in planar resonators, we believe
that the acoustical resonator could be a good candidate for the experimental
observation of transverse patterns in a planar (Fabry--Perot) cavity.
Experimental work in this direction is in progress.

\section*{Acknowledgements}

The work has been financially supported by the CICYT of the Spanish
Government, under the\ project BFM2002-04369-C04-04.

\pagebreak

{\Large Figure Captions}

\bigskip

Figure 1. Scheme of a three-element acoustic resonator. Each section is
acoustically characterized by its density an propagation velocity of sound.

\bigskip

Figure 2. Developement of stripped patterns for $\Delta_{1}=-1.6,$%
\ $\Delta_{0}=0,$ $\gamma_{0}=\gamma_{1}=5~10^{3}$ and $\mathcal{E}=2,$ as
obtained by numerical integration of Eqs. (\ref{model2})$.$ The distributions
correspond to evolution times $t=0.01$ s (a) $t=0.1$ s (b) and $t=1$ s (c).

\bigskip

Figure 3. Developement of squared patterns for $\Delta_{1}=-4,$\ $\Delta
_{0}=-8$, $\gamma_{0}=\gamma_{1}=5~10^{3}$ and $\mathcal{E}=1.2,$ as obtained
by numerical integration of Eqs. (\ref{model2})$.$ The distributions
correspond to evolution times $t=0.01$ s (a) $t=0.1$ s (b) and $t=1$ s (c).

\end{document}